\centerline{\bf A PATHWAY FROM BAYESIAN STATISTICAL ANALYSIS}
 \vskip.1cm\centerline{\bf TO SUPERSTATISTICS}

\vskip.3cm\noindent\centerline{A.M. MATHAI}
\vskip0cm\centerline{Centre for Mathematical Sciences Pala Campus}\vskip0cm\centerline{Arunapuram P.O., Pala, Kerala-686574, India and }\vskip0cm\centerline{Department of Mathematics and Statistics, McGill University, Canada, H3A2K6}
\vskip0cm\centerline{Email: mathai@math.mcgill.ca}

\vskip.3cm\centerline{and}
\vskip.3cm\centerline{H.J. HAUBOLD}
\vskip0cm\centerline{Office for Outer Space Affairs, United Nations}
\vskip0cm\centerline{P.O. Box 500, Vienna International Centre, A-1400 Vienna, Austria and}\vskip0cm\centerline{Centre for Mathematical Sciences Pala Campus}\vskip0cm\centerline{Arunapuram P.O., Pala, Kerala-686574, India}
\vskip0cm\centerline{Email: hans.haubold@unvienna.org}

\vskip.2cm\centerline{Dedicated to Professor H.M. Srivastava on the Occasion of his Seventieth Birth Anniversary}

\vskip.5cm\centerline{\bf Abstract} Superstatistics and Tsallis statistics in statistical mechanics is given an interpretation in terms of Bayesian statistical analysis. Subsequently superstatistics is extended by replacing each component of the conditional and marginal densities by Mathai's pathway model and further both components are replaced by Mathai's pathway models. This produces a wide class of mathematically and statistically interesting functions for prospective applications in statistical physics. It is pointed out that the final integral is a particular case of a general class of integrals introduced by the authors earlier. Those integrals are also connected to Kraetzel integrals in applied analysis, inverse Gaussian densities in stochastic processes, reaction rate integrals in the theory of nuclear astrophysics and Tsallis statistics in nonextensive statistical mechanics. The final results are obtained in terms of Fox's H-function. Matrix variate analogue of one significant specific case is also pointed out.


\vskip.3cm\noindent Subject Classification Numbers: 33C60, 82B31, 60E05, 62E15, 62C10

\vskip.5cm\noindent{\bf 1.\hskip.3cm Introduction}

Nonequilibrium systems in various areas of science, see Ebeling and Sokolov (2005) and Uffink (2007), exhibit spatio-temporal dynamics that is inhomogeneous and can be described by a superposition of several statistics on different scales, in short called a superstatistics approach in the physical literature, see Beck (2006). Essential for this superstatistics approach is the existence of sufficient time scale separation between two relevant dynamics within the nonequilibrium system. There must be some intensive parameter that fluctuates on a larger time scale than the typical relaxation time of the local dynamics. In a statistical thermodynamic setting such a parameter can be interpreted as a local inverse temperature of the system, but much broader interpretations are possible dependend on the nonequilibrium system under consideration. The stationary distributions of superstatistical systems, obtained by averaging over all relevant fluctuating parameters exhibit non-Gaussian behaviour with fat tails, which can be a power law, or a streched exponential. or other functional forms like Mittag-Leffler function or Fox H-function. The relevant superstatistical paremeter can be an effective parameter in a stochastic differential equation or a local variance parameter extracted from a time series. Currently, in the physical literature, superstatistics [Beck (2005)] and Tsallis statistics [Tsallis (1988)] are the two preferred models to describe nonequilibrium systems in a sense of generalyzing the well-established Boltzmann-Gibbs statistical mechanics.

In this paper we shall develop a Bayesian approach to superstatistics and Tsallis statistics in terms of the recently introduced pathway model of Mathai (2005). It will be shown that Mathai's pathway model comprises the distributions of superstatistical systems as well as Tsallis statistics in a unified manner.  

\vskip.5cm\noindent{\bf 2.\hskip.3cm Bayesian Statistical Analysis}

\vskip.3cm Let us start with the standard Bayesian analysis problem. Consider a positive real scalar random variable $x$ and a parameter $\theta$. Let the conditional density of $x$ at a given value of $\theta$ be denoted by $f(x|\theta)$ and the marginal density of $\theta$ by $g(\theta)$ respectively. For simplicity, let us take both $f(x|\theta)$ and $g(\theta)$ in the generalized gamma family of functions. Let

$$f(x|\theta)={{\delta \theta^{\gamma}}\over{\Gamma({{\gamma}\over{\delta}})}}x^{\gamma-1}{\rm e}^{-{\theta^{\delta}x^{\delta}}}\eqno(1)
$$for $\theta>0,~\delta>0,~x\ge 0,~\gamma>0$ and $f(x|\theta)=0$ elsewhere. Let

$$g(\theta)={{\delta b^{\rho/\delta}}\over{\Gamma({{\rho}\over{\delta}})}}\theta^{\rho-1}{\rm e}^{-b\theta^{\delta}}\eqno(2)
$$for $b>0,~\theta>0,~\delta>0,~\rho>0$ and $g(\theta)=0$ elsewhere. Then the unconditional density of $x$ is given by the following:

$$\eqalignno{f_x(x)&=\int_{\theta=0}^{\infty}f(x|\theta)g(\theta){\rm d}\theta={{\delta^2b^{\rho/\delta}x^{\gamma-1}}\over{\Gamma({{\gamma}
\over{\delta}})\Gamma({{\rho}\over{\delta}})}}\int_0^{\infty}\theta^{\gamma
+\rho-1}{\rm e}^{-\theta^{\delta}(b+x^{\delta})}{\rm d}\theta\cr
&={{\delta b^{\rho/\delta}\Gamma({{\gamma+\rho}\over{\delta}})}\over{\Gamma({{\gamma}
\over{\delta}})\Gamma({{\rho}\over{\delta}})}}{{x^{\gamma-1}}
\over{(b+x^{\delta})^{{\gamma+\rho}\over{\delta}}}},~~b+x^{\delta}>0\cr
&={{\delta\Gamma({{\gamma+\rho}\over{\delta}})}\over{\Gamma({{\gamma}
\over{\delta}})\Gamma({{\rho}\over{\delta}})b^{\gamma/\rho}}}x^{\gamma-1}
(1+{{x^{\delta}}\over{b}})^{-({{\gamma+\rho}\over{\delta}})}&(3)\cr}
$$for $x\ge 0,~\gamma>0,~\rho>0,~\delta>0,~b>0.$

\vskip.5cm\noindent{\bf Theorem 1.}\hskip.3cm{\it Let the conditional density of $x$ given $\theta$, that is,  $f(x|\theta)$, and the marginal density of $\theta$ be as in (1) and (2) respectively. Then the unconditional density of $x$, denoted by $f_x(x)$, is given by (3).}

\vskip.3cm In a physical system the parameter $\theta$ in (1) may represent temperature so that the density $f(x|\theta)$ may represent the production of the item $x$ at a fixed temperature $\theta$ or at a given value of $\theta$. Then the marginal density of $\theta$ in (2) may represent the temperature distribution. What is the distribution of the production of $x$ over all temperature variations or averaged over the density of $\theta$? This is the unconditional density of $x$ given in (3) for the specific densities in (1) and (2). Since a density such as the one in (2) is superimposed over the density such as the one in (1), the resulting density in (3) is called superstatistics. Various interpretations of $x$ and $\theta$ in different physical systems may be seen in the original paper on superstatistics is that of Beck and Cohen (2003). Equation (3) for $\delta=1,~\gamma=1,~b={{1}\over{q-1}},~q>1$ and ${{\gamma+\rho}\over{\delta}}={{1}\over{q-1}}$ is Tsallis statistics of non-extensive statistical mechanics, see Tsallis (1988). The Bayes' density of $\theta$ is

$$g_1(\theta|x)={{f(x|\theta)g(\theta)}\over{f_x(x)}}\eqno(4)
$$which is available from (1),(2) and (3). The Bayes' estimate of $\theta$ is the conditional expectation of $\theta$ at given value of $x$, denoted by $E(\theta|x)$, where $E$ denotes statistical expectation. This is available from the conditional density of $\theta$, given $x$, in (4). Thus, Bayes' procedure, superstatistics and Tsallis statistics are all connected together as shown in (1) to (4).

\vskip.5cm\noindent{\bf 3.\hskip.3cm Extension of Bayes' Procedure}

\vskip.3cm We can extend the discussion in (1) to (4) by using the pathway model of Mathai (2005). For example, let us replace $g(\theta)$ in (2) by a pathway density, namely,

$$P_1(\theta)={{\delta [b(\beta-1)]^{\rho/\delta}\Gamma({{\eta_1}\over{\beta-1}})}
\over{\Gamma({{\rho}\over{\delta}})\Gamma({{\eta_1}\over{\beta-1}}
-{{\rho}\over{\delta}})}}\theta^{\rho-1}[1+b(\beta-1)\theta^{\delta}]^{
-{{\eta_1}\over{\beta-1}}}\eqno(5)
$$for $\beta>1,~\delta>0,~\eta_1>0,~b>0,~\rho>0,~\theta>0,~{{\eta_1}
\over{\beta-1}}-{{\rho}\over{\delta}}>0$ and $P_1(\theta)=0$ elsewhere. When $\theta\rightarrow 1_{+}$ in (5) the model in (5) goes to $g(\theta)$ in (2) with $b$ replaced by $b\eta_1$. The model in (5) consists of three different functional forms. For $\beta<1$ we may write
$$[1+b(\beta-1)\theta^{\delta}]^{-{{\eta_1}\over{\beta-1}}}
=[1-b(1-\beta)\theta^{\delta}]^{{{\eta_1}\over{1-\beta}}}.
$$The right side remains positive in the finite range $1-b(1-\beta)\theta^{\delta}>0$ or $0<\theta<[b(1-\beta)]^{-{{1}\over{\delta}}}$. Then for $\beta<1$ the density in (5) changes to the form

$$P_2(\theta)={{\delta[b(1-\beta)]^{\rho/\delta}\Gamma({{\eta_1}
\over{1-\beta}}
+1+{{\rho}\over{\delta}})}\over{\Gamma({{\rho}\over{\delta}})\Gamma({{\eta_1}
\over{1-\beta}}+1)}}\theta^{\rho-1}[1-b(1-\beta)\theta^{\delta}]^{{{\eta_1}
\over{1-\beta}}}.\eqno(6)
$$Note that for $\beta>1$, the density in (5) stays in the generalized type-2 beta family of densities and for $\beta<1$ the density in (6) belongs to the generalized type-1 beta family of densities. When $\beta\rightarrow 1$, either from the left or from the right, $P_2(\theta)$ and $P_1(\theta)$ will go to $P_3(\theta)$, where

$$P_3(\theta)={{\delta(b\eta_1)^{\rho/\delta}}\over{\Gamma({{\rho}
\over{\delta}})}}\theta^{\rho-1}{\rm e}^{-b\eta_1\theta^{\delta}},\eqno(7)
$$for $b>0,~\eta_1>0,~\delta>0,~\rho>0,~\theta>0$, which is the generalized gamma family of densities. Thus, the parameter $\beta$ creates a path of going from the generalized type-1 beta family to generalized type-2 beta family to generalized gamma family of functions. It is shown in Mathai and Haubold (2007) that almost all statistical densities in current use in different disciplines are available from (5). In a physical system, if (7) is the stable situation then the unstable neighborhoods are covered by (5) and (6) and the paths in between, the path described by the parameter $\beta$. Note that the model in (6) is a model with the right tail cut off. The movement of $\beta$ will provide thicker or thinner-tailed distributions which will also be helpful in a model building situation where one may be looking for a thicker or thinner-tailed distribution as an appropriate fit.

\vskip.2cm It is not difficult to show that the normalizing constants in (5) and (6) reduce to the normalizing constant in (7). This can be seen by expanding the gammas with the help of Stirling's formula, which is given by

$$\Gamma(z+\alpha)\approx \sqrt{2\pi}z^{z+\alpha-{1\over2}}{\rm e}^{-z}
$$for $|z|\rightarrow\infty$ and $\alpha$ a bounded quantity. When $\beta\rightarrow 1_{-}$ we have ${{\eta_1}\over{1-\beta}}\rightarrow\infty$ and when $\beta\rightarrow 1_{+}$, ${{\eta_1}\over{\beta-1}}\rightarrow\infty$. Hence take $z={{\eta_1}\over{1-\beta}}$ or $z={{\eta_1}\over{\beta-1}}$ as the case may be, expand and simplify to see that the normalizing constants in (5) and (6) go to the constant in (7).
\vskip.2cm Instead of $g(\theta)$ of (2) if we use the extended density $P_1(\theta)$ of (5) then the unconditional density $f_x(x)$ in (3) has the following form:

$$\eqalignno{f_x(x)&=\int_{\theta=0}^{\infty}f(x|\theta)P_1(\theta){\rm d}\theta\cr
&={{\delta^2[b(\beta-1)]^{\rho/\delta}\Gamma({{\eta_1}\over{\beta-1}})}
\over{\Gamma({{\gamma}\over{\delta}})\Gamma({{\rho}\over{\delta}})
\Gamma({{\eta_1}\over{\beta-1}}-{{\rho}\over{\delta}})}}x^{\gamma-1}
\int_{\theta=0}^{\infty}\theta^{\gamma+\rho-1}[1+b(\beta-1)
\theta^{\delta}]^{-{{\eta_1}\over{\beta-1}}}{\rm e}^{-(\theta x)^{\delta}}{\rm d}\theta. &(8)\cr}
$$The integral part in (8) can be evaluated by using Mellin convolution property by observing that the integrand is of the form $\theta h_1(x\theta)h_2(\theta)$ for some functions $h_1$ and $h_2$. Let

$$h_1(\theta)=c_1\theta^{\gamma-1}{\rm e}^{-\theta^{\delta}}\hbox{  and  }h_2(\theta)=c_2\theta^{\rho-1}[1+b(\beta-1)\theta^{\delta}]^{-{{\eta_1}
\over{\beta-1}}}
$$so that $h_1$ and $h_2$ can create statistical densities for appropriate values of the normalizing constants $c_1$ and $c_2$. Therefore

$$\eqalignno{c_1c_2\int_{\theta=0}^{\infty}\theta(x\theta)^{\gamma-1}
\theta^{\rho-1}&[1+b(\beta-1)\theta^{\delta}]^{-{{\eta_1}\over{\beta-1}}}{\rm e}^{-(x\theta)^{\delta}}{\rm d}\theta\cr
&=\int_{\theta=0}^{\infty}\theta h_1(x\theta)h_2(\theta){\rm d}\theta.\cr}
$$Hence the Mellin transform of the left side is the product of the Mellin transforms of the right side with parameters $s$ for $h_1$ and $2-s$ for $h_2$. Writing in terms of statistical expectations,

$$\eqalignno{E(x_1^{s-1})&=c_1\int_0^{\infty}x_1^{\gamma-1+s-1}{\rm e}^{-x_1^{\delta}}{\rm d}x_1=c_1{{1}\over{\delta}}\Gamma({{\gamma+s-1}\over{\delta}}),
~\Re(\gamma+s-1)>0\cr
E(x_2^{1-s})&=c_2\int_0^{\infty}x_2^{1-s+\rho-1}[1+b(\beta-1)
x_2^{\delta}]^{-{{\eta_1}\over{\beta-1}}}{\rm d}x_2\cr
&=c_2{{1}\over{\delta}}{{\Gamma({{\rho-s+1}\over{\delta}})
\Gamma({{\eta_1}\over{\beta-1}}-{{\rho-s+1}\over{\delta}})}
\over{[b(\beta-1)]^{{{\rho-s+1}\over{\delta}}}\Gamma({{\eta_1}
\over{\beta-1}})}}\cr}
$$for $\Re(\rho-s+1)>0,~\Re({{\eta_1}\over{\beta-1}}-{{\rho-s+1}\over{\delta}})>0$. Therefore the inverse Mellin transform is the following:

$$\eqalignno{c_1c_2&\int_{\theta=0}^{\infty}\theta(x\theta)^{\gamma-1}
\theta^{\rho-1}[1+b(\beta-1)\theta^{\delta}]^{-{{\eta_1}\over{\beta-1}}}
{\rm e}^{-(\theta x)^{\delta}}{\rm d}\theta\cr
&=c_1c_2{{1}\over{2\pi i}}\int_{c-i\infty}^{c+i\infty}{{1}\over{\delta^2\Gamma({{\eta_1}
\over{\beta-1}})[b(\beta-1)]^{{\rho+1}\over{\delta}}}}\cr
&\times \Gamma({{\gamma-1}\over{\delta}}+{{s}\over{\delta}})\Gamma({{\rho+1}
\over{\delta}}-{{s}\over{\delta}})\Gamma({{\eta_1}\over{\beta-1}}-{{\rho+1}
\over{\delta}}+{{s}\over{\delta}})[{{1}\over{[b(\beta-1)]^{{1}
\over{\delta}}}}]^{-s}{\rm d}s\cr
&=c_1c_2{{1}\over{\delta^2\Gamma({{\eta_1}\over{\beta-1}})
[b(\beta-1)]^{{\rho+1}\over{\delta}}}}H_{1,2}^{2,1}\left[{{x}
\over{[b(\beta-1)]^{{1}\over{\delta}}}}\big\vert_{({{\gamma-1}\over{\delta}},{{1}\over{\delta}}),({{\eta_1}\over{\beta-1}}-{{\rho+1}\over{\delta}},{{1}\over{\delta}})}^{(1-{{\rho+1}\over{\delta}},{{1}\over{\delta}})}\right]\cr}
$$where H is the H-function, see for example, Mathai, Saxena, and Haubold (2010) and Srivastava, Gupta, and Goyal (1982).

\vskip.5cm\noindent{\bf Theorem 2.\hskip.3cm}{\it When the conditional density of $x$ given $\theta$ is of the form $f(x|\theta)$ in (1), when the marginal density of $\theta$ is of the form of $P_1(\theta)$ in (5) then the unconditional density of $x$, denoted by $f_x(x)$ is given by

$$\eqalignno{f_x(x)&={{1}\over{[b(\beta-1)]^{{1}\over{\delta}}
\Gamma({{\gamma}\over{\delta}})\Gamma({{\rho}\over{\delta}})
\Gamma({{\eta_1}\over{\beta-1}}-{{\rho}\over{\delta}})}}\cr
&\times H_{1,2}^{2,1}\left[{{x}\over{[b(\beta-1)]^{{1}\over{\delta}}}}
\big\vert_{({{\gamma-1}\over{\delta}},{{1}\over{\delta}}),
({{\eta_1}\over{\beta-1}}-{{\rho+1}\over{\delta}},
{{1}\over{\delta}})}^{(1-{{\rho+1}\over{\delta}},
{{1}\over{\delta}})}\right]&(9)\cr}
$$for $b>0,~\beta>1,~\gamma>1,~\rho>0,~\delta>0,~{{\eta_1}\over{\beta-1}}
-{{\rho+1}\over{\delta}}>0$.}

\vskip.5cm\noindent{\bf 4.\hskip.3cm Pathway Extension of the Conditional Density}

\vskip.3cm Let us consider an extended form of the conditional density $f(x|\theta)$ and the original form of the marginal density as in (2). Let the extended conditional density be of the pathway model type-2, given by,
$$f(x|\theta)={{\delta[a(\alpha-1)]^{\gamma/\delta}\Gamma({{\eta}
\over{\alpha-1}})}\over{\Gamma({{\gamma}\over{\delta}})\Gamma({{\eta}
\over{\alpha-1}}-{{\gamma}\over{\delta}})}}\theta^{\gamma}
x^{\gamma-1}[1+a(\alpha-1)(\theta x)^{\delta}]^{-{{\eta}\over{\alpha-1}}}\eqno(10)
$$for $\alpha>1,~\delta>0,~\eta>0,~\gamma>0,~\theta>0,~x>0$ and $f(x|\theta)=0$ elsewhere. Let the marginal density of $\theta$ be as in (2). Then the unconditional density of $x$, again denoted by $f_x(x)$, is given by the following:

$$\eqalignno{f_x(x)&=\int_{\theta=0}^{\infty}f(x|\theta)g(\theta){\rm d}\theta\cr
&={{\delta^2[a(\alpha-1)]^{\gamma/\delta}b^{\rho/\delta}}
\over{\Gamma({{\rho}\over{\delta}})\Gamma({{\gamma}\over{\delta}})
\Gamma({{\eta}\over{\alpha-1}}-{{\gamma}\over{\delta}})}}x^{\gamma-1}\cr
&\times \int_{\theta=0}^{\infty}\theta^{\gamma+\rho-1}[1+a(\alpha-1)
(x\theta)^{\delta}]^{-{{\eta}\over{\alpha-1}}}{\rm e}^{-b\theta^{\delta}}{\rm d}\theta.\cr}
$$The integral part can be written as follows:

$$\eqalignno{c_1c_2&\int_{\theta=0}^{\infty}(x\theta)^{\delta}]^{-{{\eta}
\over{\alpha-1}}}{\rm e}^{-b\theta^{\delta}}{\rm d}\theta\cr
&=\int_{\theta=0}^{\infty}\theta h_1(x\theta)h_2(\theta){\rm d}\theta\cr
\noalign{\hbox{where}}
h_1(x_1)&=c_1x_1^{\gamma-1}[1+a(\alpha-1)x_1^{\delta}]^{-{{\eta}
\over{\alpha-1}}}\cr
\noalign{\hbox{and}}
h_2(x_2)&=c_2 x_2^{\rho-1}{\rm e}^{-bx_2^{\delta}}.\cr}
$$Then $c_1$ and $c_2$ can act as normalizing constants so that $h_1$ and $h_2$ are statistical densities. Then, writing the Mellin transform in terms of expected values, we have the following:

$$\eqalignno{E(x_1^{s-1})&=c_1\int_{x_1=0}^{\infty}x_1^{\gamma+s-2}
[1+a(\alpha-1)x_1^{\delta}]^{-{{\eta}\over{\alpha-1}}}{\rm d}x_1\cr
&=c_1{{1}\over{\delta[a(\alpha-1)]^{{\gamma+s-1}\over{\delta}}}}
{{\Gamma({{\gamma+s-1}\over{\delta}})\Gamma({{\eta}\over{\alpha-1}}
-{{\gamma+s-1}\over{\delta}})}\over{\Gamma({{\eta}\over{\alpha-1}})}}\cr
\noalign{\hbox{for  $\Re(\gamma+s-1)>0,~\Re({{\eta}\over{\alpha-1}}
-{{\gamma+s-1}\over{\delta}})>0.$}}
E(x_2^{1-s})&=c_2\int_{x_2=0}^{\infty}x_2^{\rho+s-2}{\rm e}^{-bx_2^{\delta}}{\rm d}x_2\cr
&=c_2{{\Gamma({{\rho+s-1}\over{\delta}})}\over{\delta b^{{\rho+s-1}\over{\delta}}}}\hbox{  for  }\Re(\rho+s-1)>0.\cr}
$$Hence the inverse Mellin transform of $E(x_1^{s-1})E(x_2^{1-s})$ is given by
$$\eqalignno{{{1}\over{2\pi i}}\int_{c-i\infty}^{c+i\infty}&{{1}\over{\delta^2
[a(\alpha-1)]^{{\gamma-1}\over{\delta}}b^{{\rho-1}
\over{\delta}}\Gamma({{\eta}\over{\alpha-1}})}}\cr
&\times \Gamma({{\gamma-1}\over{\delta}}+{{s}\over{\delta}})
\Gamma({{\rho-1}\over{\delta}}
+{{s}\over{\delta}})\Gamma({{\eta}\over{\alpha-1}}-{{\gamma-1}\over{\delta}}
-{{s}\over{\delta}})[a(\alpha-1)]^{-{{s}\over{\delta}}}b^{-{{s}\over{\delta}}}
x^{-s}{\rm d}s\cr
&={{1}\over{\delta^2[a(\alpha-1)]^{{\gamma-1}\over{\delta}}b^{{\rho-1}
\over{\delta}}\Gamma({{\eta}\over{\alpha-1}})}}\cr
&\times H_{1,2}^{2,1}\left[(a(\alpha-1))^{{1}\over{\delta}}b^{{1}
\over{\delta}}x\big\vert_{({{\gamma-1}\over{\delta}},{{1}
\over{\delta}}),({{\rho-1}\over{\delta}},{{1}
\over{\delta}})}^{(1+{{\gamma-1}\over{\delta}}-{{\eta}\over{\alpha-1}},{{1}
\over{\delta}})}\right]&(11)\cr}
$$
\vskip.3cm\noindent{\bf Theorem 3.\hskip.3cm}{\it Let the conditional density of $x$ at given $\theta$ be given by (10) and the marginal density of $\theta$ be given by (2). Then the unconditional density of $x$, denoted by $f_x(x)$, is given by

$$\eqalignno{f_x(x)&={{[ab(\alpha-1)]^{1/\delta}}
\over{\Gamma({{\rho}\over{\delta}})\Gamma({{\gamma}\over{\delta}})
\Gamma({{\eta}\over{\alpha-1}})\Gamma({{\eta}\over{\alpha-1}}
-{{\gamma}\over{\delta}})}}\cr
&\times H_{1,2}^{2,1}\left[[ab(\alpha-1)]^{{1}\over{\delta}}x
\big\vert_{({{\gamma-1}\over{\delta}},{{1}\over{\delta}}),
({{\rho-1}\over{\delta}},{{1}\over{\delta}})}^{(1+{{\gamma-1}\over{\delta}}
-{{\eta}\over{\alpha-1}},{{1}\over{\delta}})}\right]&(12)\cr}
$$for $\gamma-1>0,~{{\eta}\over{\alpha-1}}-{{\gamma-1}\over{\delta}}>0,~\rho-1>0$.}

\vskip.3cm\noindent{\bf 5.\hskip.3cm General Pathway Model for Unconditional Densities}

\vskip.3cm Now we consider a more general case where $f(x|\theta)$ is given in (10) and $g(\theta)$ is given in (5). Then what will be the unconditional density of $x$? From the procedure so far, it is clear that the unconditional density of $x$, denoted by $f_x(x)$,is given by the following:
$$\eqalignno{f_x(x)&={{\delta[b(\beta-1)]^{\rho/\delta}
\Gamma({{\eta_1}\over{\beta-1}})}\over{\Gamma({{\rho}\over{\delta}})
\Gamma({{\eta_1}\over{\beta-1}}-{{\rho}\over{\delta}})}}
{{\delta[a(\alpha-1)]^{\gamma/\delta}\Gamma({{\eta}\over{\alpha-1}})}
\over{\Gamma({{\gamma}\over{\delta}})\Gamma({{\eta}\over{\alpha-1}}
-{{\gamma}\over{\delta}})}}\cr
&\times x^{\gamma-1}\int_{\theta=0}^{\infty}\theta^{\gamma}[1+a(\alpha-1)\theta^{\delta}
x^{\delta}]^{-{{\eta}\over{\alpha-1}}}\cr
&\times \theta^{\rho-1}[1+b(\beta-1)\theta^{\delta}]^{-{{\eta_1}\over{\beta-1}}}{\rm d}\theta &(13)\cr}
$$for $\beta>1,~\alpha>1,~\delta>0,~\eta>0,~\eta_1>0,~b>0,~a>0,~\rho>0$, $\theta>0,~x>0, ~{{\eta}\over{\alpha-1}}-{{\gamma}\over{\delta}}>0,~{{\eta_1}\over{\beta-1}}
-{{\rho}\over{\delta}}>0$. Consider the integral part.

$$\eqalignno{\int_{\theta=0}^{\infty}\theta&(x\theta)^{\gamma-1}\theta^{\rho-1}[1+a(\alpha-1)(x\theta)^{\delta}]^{-{{\eta}\over{\alpha-1}}}[1+b(\beta-1)\theta^{\delta}]^{-{{\eta_1}\over{\beta-1}}}{\rm d}\theta\cr
&=\int_{\theta=0}^{\infty}\theta h_1(x\theta)h_2(\theta){\rm d}\theta\cr
\noalign{\hbox{where}}
h_1(x_1)&=c_1x_1^{\gamma-1}[1+a(\alpha-1)x_1^{\delta}]^{-{{\eta}\over{\alpha-1}}}\cr
\noalign{\hbox{and}}
h_2(x_2)&=c_2x_2^{\rho-1}[1+b(\beta-1)x_2^{\delta}]^{-{{\eta_1}\over{\beta-1}}}.\cr}
$$When $c_1$ and $c_2$ are normalizing constants then $h_1(x_1)$ and $h_2(x_2)$ can act as statistical densities.

$$\eqalignno{E(x_1^{sd-1})&=c_1\int_0^{\infty}x_1^{\gamma+s-2}[1+a(\alpha-1)x_1^{\delta}]^{-{{\eta}\over{\alpha-1}}}{\rm d}x_1\cr
&=c_1{{\Gamma({{\gamma+s-1}\over{\delta}})\Gamma({{\eta}\over{\alpha-1}}-{{\gamma+s-1}\over{\delta}})}\over{\delta[a(\alpha-1)]^{{\gamma+s-1}\over{\delta}}\Gamma({{\eta}\over{\alpha-1}})}}\cr
&\hbox{for  }\Re(\gamma+s-1)>0,~\Re({{\eta}\over{\alpha-1}}-{{\gamma+s-1}\over{\delta}})>0,~a>0,~\alpha>1,~\delta>0,\eta>0.\cr
E(x_2^{1-s})&=c_2\int_0^{\infty}x_2^{\rho-s}[1+b(\beta-1)x_2^{\delta}]^{-{{\eta_1}\over{\beta-1}}}{\rm d}x_2\cr
&=c_2{{\Gamma({{\rho-s+1}\over{\delta}})\Gamma({{\eta_1}\over{\beta-1}}-{{\rho-s+1}\over{\delta}})}\over{\delta[b(\beta-1)]^{{\rho-s+1}\over{\delta}}\Gamma({{\eta_1}\over{\beta-1}})}}\cr
&\hbox{for  }\Re(\rho-s+1)>0,~\Re({{\eta_1}\over{\beta-1}}-{{\rho-s+1}\over{\delta}})>0,~\delta>0,~b>0,~\beta>1,~\eta_1>0.\cr}
$$Then the integral is available from the inverse Mellin transform which can again be written in terms of a H-function.

\vskip.5cm\noindent{\bf Theorem 4.\hskip.3cm}{\it Let the conditional density of $x$ given $\theta$ be given in (10) and the marginal density of $\theta$ be given in (5). Then the unconditional density, denoted by $f_x(x)$, is given by
$$\eqalignno{f_x(x)&=\left[{{a(\alpha-1)}\over{b(\beta-1)}}\right]^{{1}\over{\delta}}
{{1}\over{\Gamma({{\rho}\over{\delta}})\Gamma({{\gamma}\over{\delta}})
\Gamma({{\eta_1}\over{\beta-1}}-{{\rho}\over{\delta}})
\Gamma({{\eta}\over{\alpha-1}}-{{\gamma}\over{\delta}})}}\cr
&\times H_{2,2}^{2,2}\left[\left[{{a(\alpha-1)}\over{b(\beta-1)}}\right]^{{1}\over{\delta}}x
\big\vert_{({{\gamma-1}\over{\delta}},{{1}\over{\delta}}),
({{\eta_1}\over{\beta-1}}-{{\rho+1}\over{\delta}},{{1}\over{\delta}})}^{(1
+{{\gamma-1}\over{\delta}}-{{\eta}\over{\alpha-1}},{{1}\over{\delta}}),
(1-{{\rho+1}\over{\delta}},{{1}\over{\delta}})}\right]&(14)\cr}
$$for $a>0,~b>0,~\alpha>1,~\beta>1,~\gamma>1$, $\rho>0,~\delta>0,~x>0,~{{\eta_1}\over{\beta-1}}-{{\rho}\over{\delta}}>0,
~{{\eta}\over{\alpha-1}}-{{\gamma}\over{\delta}}>0$.}

\vskip.5cm\noindent{\bf Remark 1.}\hskip.3cm The integrand in this general case, giving rise to Theorem 4,  is in the form of a product of two pathway models. This, in fact is a special case of a versatile integral considered in Mathai (2007). This will be briefly discussed in the next section.

\vskip.5cm\noindent{\bf 6.\hskip.3cm A Versatile Integral}

\vskip.3cm One form of the versatile integral is the following:

$$f(z_2|z_1)=\int_0^{\infty}x^{\gamma-1}[1+z_1^{\delta}(\alpha-1)
x^{\delta}]^{-{{1}\over{\alpha-1}}}[1+z_2^{\rho}(\beta-1)
x^{-\rho}]^{-{{1}\over{\beta-1}}}{\rm d}x\eqno(15)
$$where the parameters $\alpha$ and $\beta$ can be independently less than one, greater than one and going to one. Also $\delta$ and $\rho$ can be independently positive or negative, all are assumed to be real quantities for convenience. Thus the integral in (15)  covers a large spectrum of integrals. The integrand in (15) is nothing but a product of two pathway models in the real scalar case. When $\alpha\rightarrow 1$ and $\beta\rightarrow 1$ we have several important integrals in different fields: (1): Kr\"atzel integral in applied analysis is a particular case in this situation with $\delta=1$ and $\rho=1$. (2): When $\rho={1\over2}$ and $\delta=1$  we have the reaction rate probability integral in the theory of nuclear astrophysics discussed in a series of papers, see for example, Mathai and Haubold (1988). (3): When $\rho=1$ one has the inverse Gaussian density, which appears very frequently in time series, stochastic processes and statistical distribution theory. Generalizations of all these practical situations are given by (15) for various values of $\alpha,~\beta,~\delta,~\rho$.

\vskip.5cm\noindent{\bf 7.\hskip.3cm Matrix-variate Analogues}
\vskip.3cm All cases of (1) to (15) cannot be given matrix-variate analogues due to restrictions on Jacobians of matrix transformations, see for example Mathai (1997). One situation is straightforward. Consider  real positive definite $p\times p$ matrices $X$ and $Y$ having densities belonging to the matrix-variate gamma families. Let the conditional density of $Y$, given $X$, be given by

$$\eqalignno{f(Y|X)&={{|X|^{{p+1}\over2}}\over{\Gamma_p(\alpha)}}
|XY|^{\alpha-{{p+1}\over2}}{\rm e}^{-{\rm tr}(XY)},~X>0,~Y>0\cr
\noalign{\hbox{and}}
g(X)&={{|B|^{\beta}}\over{\Gamma_p(\beta)}}|X|^{\beta-{{p+1}\over2}}{\rm e}^{-{\rm tr}(BX)},~B>0\cr}
$$where $|(\cdot)|$ denotes the determinant of $(\cdot)$, ${\rm tr}(\cdot)$ the trace of $(\cdot)$ and the real matrix-variate gamma function is given as

$$\Gamma_p(\alpha)=\pi^{{p(p-1)}\over4}\Gamma(\alpha)\Gamma(\alpha-{1\over2})
...\Gamma(\alpha-{{p-1}\over2}),~\Re(\alpha)>{{p-1}\over2}.
$$The standard notation $A>0$ is used for the matrix $A$ being real symmetric, and further, positive definite. Then the unconditional density of $Y$, denoted by $f_Y(Y)$, is given by the following:

$$\eqalignno{f_Y(Y)&=\int_Xf(Y|X)g(X){\rm d}X\cr
&={{|B|^{\beta}}\over{\Gamma_p(\alpha)\Gamma_p(\beta)}}
\int_X|Y|^{\alpha-{{p+1}\over2}}|X|^{\alpha+\beta-{{p+1}\over2}}{\rm e}^{-{\rm tr}[X(B+Y)]}{\rm d}X\cr
&={{|B|^{\beta}}\over{\Gamma_p(\alpha)\Gamma_p(\beta)}}
|Y|^{\alpha-{{p+1}\over2}}|B+Y|^{-(\alpha+\beta)}\Gamma_p(\alpha+\beta)\cr
&={{\Gamma_p(\alpha+\beta)}\over{|B|^{\alpha}\Gamma_p(\alpha)\Gamma_p(\beta)}}
|Y|^{\alpha-{{p+1}\over2}}|I+B^{-1}Y|^{-(\alpha+\beta)},~Y>0,~B>0.&(16)\cr}
$$The transformation used in the integration is $U=(B+Y)^{1\over2}X(B+Y)^{1\over2}\Rightarrow {\rm d}X=|B+Y|^{-{{p+1}\over2}}{\rm d}U$ where, for example, ${\rm d}U$ denotes the wedge product of the $p(p+1)/2$ differentials in the elements of $U$ and $(\cdot)^{1\over2}$ denotes the positive definite square root of the positive definite matrix $(\cdot)$. From (16) it is clear that the unconditional density of $Y$ is a matrix-variate type-2 beta density. This is the matrix-variate analogue of the superstatistics in the case of conditional and marginal densities belonging to the gamma type densities.

\vskip.3cm\noindent\centerline{\bf Acknowledgement}

\vskip.3cm The authors would like to thank the Department of Science and Technology, Government of India for the financial assistance under project No. SR/S4/MS:287/05 which made the collaboration possible.

\vskip.2cm
\noindent\centerline{\bf References}

\vskip.2cm \noindent Beck, C. and Cohen, E.G.D. (2003): Superstatistics, {\it Physica A}, {\bf 322}, 267-275.

\vskip.2cm \noindent \hskip.5cm C. Beck, Stretched exponentials from superstatistics, {\it Physica A}, {\bf 365} (2006), 56-101;
\vskip.2cm \noindent \hskip.5cm C. Beck, in R. Klages, G. Radons, and I.M. Sokolov: {\it Anomalous Transport: Foundations} \par {\it and Applications}, Wiley-VCH, Weinheim, 2008, pp. 433-457;
\vskip.2cm \noindent \hskip.5cm C. Beck, Recent developments in superstatistics, {\it Brazilian Journal of Physics}, {\bf 39} (2009), 357-363;
\vskip.2cm \noindent \hskip.5cm C. Beck, Generalized statistical mechanics for superstatistical systems, arXiv.org/1007.0903\par [cond-mat.stat-mech], 6 July 2010.

\vskip.2cm \noindent Ebeling, W. and Sokolov, M. (2005): {\it Statistical Thermodynamics and Stochastic Theory of Nonequilibrium Systems}, World Scientific, Singapore.

\vskip.2cm \noindent Mathai, A.M. (1997): {\it Jacobians of Matrix Transformations and Functions of Matrix Argument}, World Scientific Publishers, New York.

\vskip.2cm \noindent Mathai, A.M. (2005): A pathway to matrix-variate gamma and normal densities. {\it Linear Algebra and Its Applications}, {\bf 396}, 317-328.

\vskip.2cm \noindent Mathai, A.M. and Haubold, H.J. (1988): {\it Modern Problems in Nuclear and Neutrino Astrophysics}, Akademie-Verlag, Berlin.

\vskip.2cm \noindent Mathai, A.M. and Haubold, H.J. (2007): Pathway model, superstatistics, Tsallis statistics and a generalized measure of entropy, {\it Physica A}, {\bf 375}, 110-122;
\vskip.2cm \noindent \hskip.5cm A.M. Mathai and H.J. Haubold, Mittag-Leffler functions to pathway model to Tsallis statistics, {\it Integral} \par {\it Transforms and Special Functions}, {\bf 21} (2010), 867-875;
\vskip.2cm \noindent \hskip.5cm A.M. Mathai, H.J. Haubold, and C. Tsallis, Pathway model and nonextensive statistical mechanics, \par arXiv.org/abs/1010.4597 [cond-mat.stat-mech] 21 October 2010.

\vskip.2cm \noindent Mathai, A.M., Saxena, R.K., and Haubold, H.J. (2010): {\it The H-function: Theory and Applications}, Springer, New York.

\vskip.2cm \noindent Srivastava, H.M., Gupta, K.C., and Goyal S.P. (1982): {\it The H-functions of One and Two Variables With Applications}, South Asian Publishers, New Delhi.

\vskip.2cm \noindent Tsallis, C. (1988): Possible generalization of Boltzmann-Gibbs statistics, {\it Journal of Statistical Physics}, {\bf 52}, 479-487;
\vskip.2cm \noindent \hskip.5cm C. Tsallis, What should statistical mechanics satisfy to reflect nature?, {\it Physica D}, {\bf 193} (2004), 3-34;
\vskip.2cm \noindent \hskip.5cm C. Tsallis, Nonadditive entropy and nonextensive statistical mechanics - An overview after 20 years,\par {\it Brazilian Journal of Physics}, {\bf 39} (2009), 337-356;
\vskip.2cm \noindent \hskip.5cm C. Tsallis, {\it Introduction to Nonextensive Statistical Mechanics: Approaching a Complex World}, Springer,\par New York, 2009.

\vskip.2cm \noindent Uffink, J. (2007): Compendium of the foundations of classical statistical physics, in {\it Philosophy of Physics, Part B}, Eds. J. Butterfield and J. Earman, Elsevier, Amsterdam, pp. 923-1074.

\bye